\documentclass[preprint]{aastex}
\usepackage{emulateapj5}
\usepackage{apjfonts}
\usepackage{natbib}

\input psfig.tex

\def\percm2{cm$^{-2}$}

\def\lax{{$\mathrel{\hbox{\rlap{\hbox{\lower4pt\hbox{$\sim$}}}\hbox{$<$}}}$}}
\def\gax{{$\mathrel{\hbox{\rlap{\hbox{\lower4pt\hbox{$\sim$}}}\hbox{$>$}}}$}}

\slugcomment{Draft version January 1, 2004}
\lefthead{Pandey, G. et al. }
\righthead{Abundances of  neutron-capture elements in hot extreme-Helium Stars}

\begin{document}

\title{Abundances of neutron-capture elements in the Hot Extreme-Helium Stars
 V1920 Cygni and HD\,124448\footnotemark[1]}

\footnotetext[1]{Based on observations obtained with the NASA/ESA 
{\it Hubble Space Telescope}, which is operated by the Association of 
Universities for Research in Astronomy, Inc. (AURA) under NASA contract 
NAS 5-26555 }

\author{
Gajendra Pandey\altaffilmark{2,3}, David L. Lambert\altaffilmark{2},
N. Kameswara Rao\altaffilmark{3}, C. Simon Jeffery\altaffilmark{4}\\ 
 }
\altaffiltext{2}{Department of Astronomy; University of
Texas; Austin, TX 78712-1083; dll@astro.as.utexas.edu}

\altaffiltext{3}{Indian Institute of Astrophysics;
Bangalore,  560034 India; pandey@iiap.res.in, nkrao@iiap.res.in}

\altaffiltext{4}{Armagh Observatory; College Hill,
Armagh, BT61 9DG, UK; csj@star.arm.ac.uk}

\setcounter{footnote}{4}
\begin{abstract}

Analysis of  $HST$ STIS ultraviolet spectra
of two hot extreme helium stars (EHes): V1920\,Cyg and HD\,124448
provide the first measurements of abundances of neutron-capture
elements for EHes. Although the two stars have similar abundances for elements
up through the iron-group, they differ strikingly in their abundances
of heavier elements: V1920\,Cyg is enriched by a factor of 30
in light neutron-capture elements (Y/Fe, Zr/Fe) relative to HD\,124448.
These differences in abundances 
of neutron-capture elements 
among EHes mirrors that exhibited by the 
R\,CrB stars, and is evidence supporting the view that there is an evolutionary 
connection between these two groups of hydrogen-deficient stars. 
Also, the abundances of Y and Zr in V1920\,Cyg provide
evidence that at least one EHe star went through a $s$-process synthesis episode in
its earlier evolution.

\end{abstract}


\section{Introduction}                   

Extreme helium stars (EHes) are carbon-rich B- and A-type supergiants in which
surface hydrogen is merely a trace element (Jeffery et al. 1987). R\,CrB stars
are similarly hydrogen-poor carbon-rich F- and G-type supergiants characterised
by steep and irregular declines in visual brightness (Asplund et al. 2000).
Understanding the origins of these rare luminous H-deficient
stars remains a challenge. 
Detailed analyses of the star's chemical
compositions hold clues to their origins. A significant lacuna presently
exists: the abundances of neutron-capture elements in EHes
are unknown. Here, we provide and comment on the first estimates of these
abundances for a pair of EHes. 

Two scenarios are contenders to account for the
EHes and the R\,CrB stars: the 
merger of a He white dwarf with a C-O white dwarf (Webbink 1984), and a 
final shell flash in a post-AGB star on the white dwarf cooling 
track (Iben et al. 1983).
A final shell flash appears most likely responsible for the  remarkable
stars FG Sge (Langer, Kraft \& Anderson 1974; Gonzalez et al. 1998) and 
V4334 Sgr (Sakurai's object $-$ Duerbeck \& Benetti 1996; Asplund et al. 1997)
with R\,CrB-like light curves, high overabundances of the neutron-capture elements,
and probable or certain H-deficiency. White dwarf mergers appear
to account
for the compositions of EHes and (probably) the R\,CrBs (Pandey et al.
2001; Saio \& Jeffery 2002; Asplund et al. 2000).
The $s$-process abundances in EHes are crucial clues because,
in the merger model, enrichment of neutron-capture elements is not expected
unless some $s$-processing occurs during the merger (Pandey et al. 2001).


For  hot EHes ($T_{\rm eff} > 14000$ K), the 
lighter neutron-capture elements (Sr, Y, Zr) and the heavier elements
(Ba and the lanthanides) are undetectable in optical spectra because
ionization equilibrium ensures that the dominant ion is $X^{2+}$, whose
strongest lines are principally in the ultraviolet. 
Fortunately, the EHes have appreciable ultraviolet flux and
are observable at high-spectral resolution with the $Hubble$ $Space$ 
$Telescope's$ Space Telescope Imaging Spectrograph.
Here, we report abundance  analyses for a pair of
hot EHes of similar temperature and gravity and
with almost identical compositions except for
abundances of Y and Zr. 

\section{Observations}        

{\it Ultraviolet:} V1920\,Cyg (aka HD\,225642 and LS II +33 5) and HD\,124448 
were observed (program ID: GO 9417) on 2002 October 18 and 2003 July 21, 
respectively, with the $HST$'s STIS Near-UV/MAMA, using the E230M grating
and 0.$''$2 $\times$ 0.$''$06 aperture, which provides a
resolving power ($\lambda/\Delta\lambda$) of 30,000. 
Two spectra for each star were obtained: V1920\,Cyg (Data Sets O6MB06010 and 
O6MB06020 with exposure times 1844 s and 2945 s, respectively) and 
HD\,124448 (Data Sets O6MB02010 and O6MB02020 with exposure times 1977 s and
3054 s, respectively). The  spectrum covered the wavelength range from 
1840\AA\ to 2670\AA. 
Since the absorption profiles are broad for V1920\,Cyg (projected rotational
velocity $v\sin i$ $\sim$ 40 km s$^{-1}$; Jeffery et al. 1998)
and HD\,124448 ($v\sin i$ $\sim$ 20 km s$^{-1}$; Sch\"{o}nberner \& Wolf 1974),
the coadded spectra from two exposures were rebinned to a lower resolution to 
improve the signal-to-noise (S/N) ratio, which is about 100 at 2500\AA. 
The resulting spectra 
have a resolving power of 7500 (V1920\,Cyg) and 15000 (HD\,124448). 

{\it Optical:} A high-resolution optical spectrum of V1920\,Cyg was obtained
on 1996 July 25 at the W. J. McDonald Observatory's 2.7-m
telescope with the coud\'{e} cross-dispersed echelle 
spectrograph (Tull et al. 1995) at a resolving power of 60,000.
The observing procedure, the detector, and the wavelength coverage
are  as described in Pandey et al. (2001). 


\section{Abundance analysis}

Model atmospheres from the code STERNE and synthetic spectra computed 
with the Belfast LTE code SPECTRUM are combined in the 
analysis (Jeffery, Woolf, \& Pollacco 2001). 
Input parameters for STERNE including the composition were taken from
previous abundance analyses. 
These  parameters are the effective temperature
$T_{\rm eff}$ = 16180$\pm$500 K, the surface gravity
log $g$ = 2.00$\pm$0.25 (cgs units), the microturbulent velocity
$\xi$ = 15$\pm$5 km s$^{-1}$, and the abundance ratio C/He = 1\% by number 
of atoms for V1920\,Cyg (Jeffery et al. 1998), and  
$T_{\rm eff}$ = 15500$\pm$800 K, log $g$ = 2.10$\pm$0.20 (cgs units), 
$\xi$ = 10 km s$^{-1}$, and C/He = 1\% for 
HD\,124448 (Sch\"{o}nberner \& Wolf 1974; Heber 1983). Derived abundances are
sufficiently close to the input values that iteration is unnecessary.
The continuous opacity is dominated by photoionization  of neutral helium
and electron scattering with electrons supplied by helium. In this situation, 
the He\,{\sc i} line strengths are insensitive to the abundances of the trace
elements, and to $T_{\rm eff}$, but are sensitive to gravity because the
strong lines have Stark-broadened wings. The adopted models  satisfactorily
reproduce the optical and ultraviolet He\,{\sc i} line profiles (V1920\,Cyg).
The derived abundances (Table 1) from spectum syntheses
are given as $\log\epsilon$(X) and normalized 
with respect to total mass where $\log\Sigma\mu_X\epsilon$(X) = 12.15 with 
$\mu$ as the atomic weight. 

\psfig{file=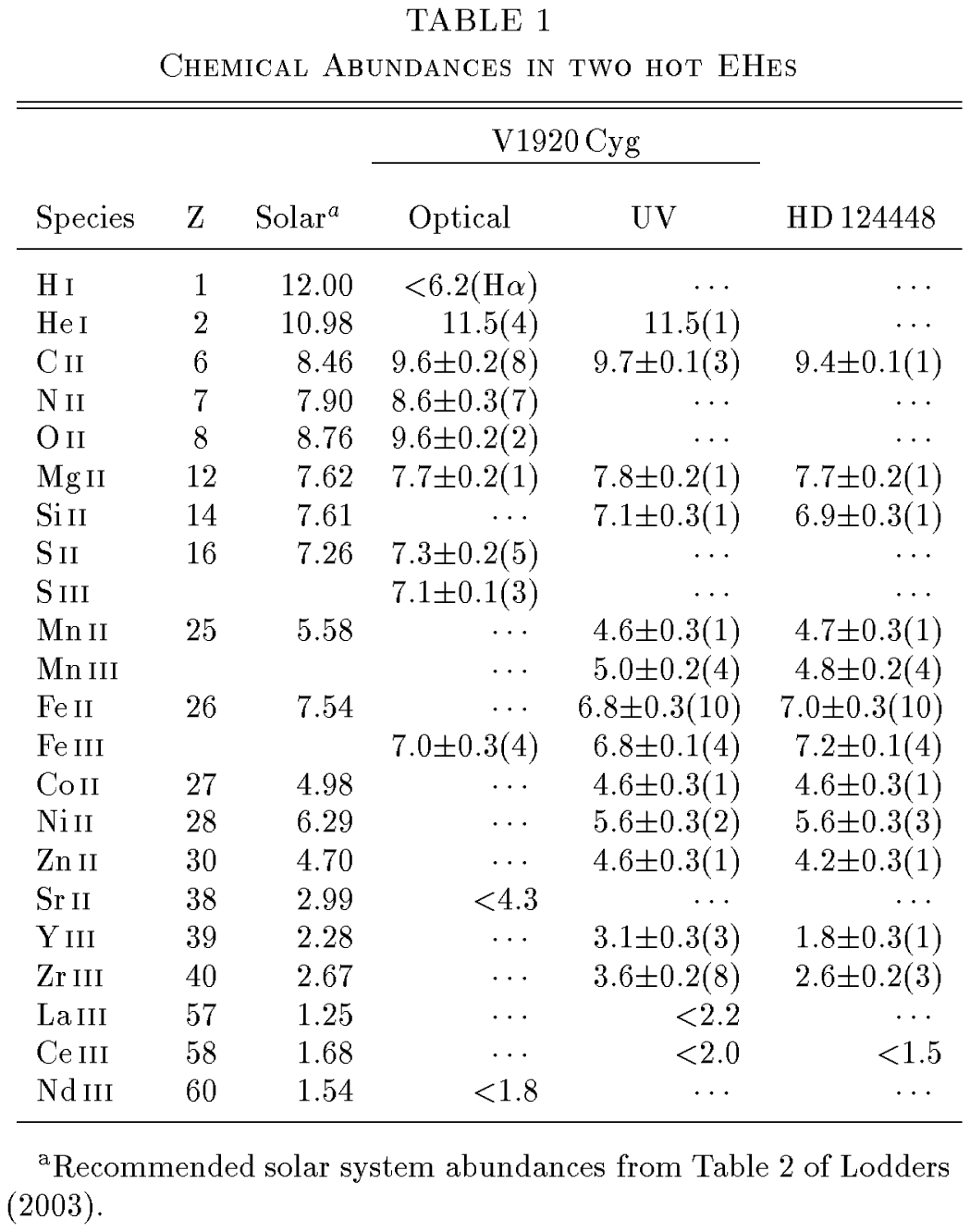,width=9.5cm,angle=0}

The adopted $gf$-values  were
taken from the NIST database\footnote{http://physics.nist.gov/cgi-bin/AtData/lines\_form} 
for H, He, Mg, Si, S, Mn\,{\sc ii}, Co, Ni, Zn, and Sr, 
Wiese, Fuhr \& Deters (1996) for C, N, and O, and  
Kurucz's database\footnote{http://kurucz.harvard.edu} for Mn\,{\sc iii} and 
Fe\,{\sc iii}. For Zr\,{\sc iii}, three sets of  theoretical $gf$-values are 
in good agreement:
Redfors (1991), Reader \& Acquista (1997), and
Charro, L\'{o}pez-Ayuso \& Mart\'{i}n (1999).
We adopt Redfors (1991) $gf$-values with the
suggested correction by Sikstr\"{o}m et al. (1999) for Y\,{\sc iii} and
Zr\,{\sc iii}; the estimated uncertainty in the $gf$-values is reported to be 
within 10\%. The $gf$-values for La\,{\sc iii}, Ce\,{\sc iii}, and 
Nd\,{\sc iii} are from 
DREAM database\footnote {ftp://mail.umh.ac.be/pub/ftp\_astro/dream/LaIII}, 
Bi\'{e}mont, Quinet \& Ryabchikova (2002), and Zhang et al. (2002), 
respectively.
The estimated abundance uncertainty is followed in brackets by the number of 
useful lines.  
The uncertainty is the combined uncertainty from the estimated errors in the
atmospheric parameters or, if larger, the line-to-line scatter in the 
abundances. Note that for the ultraviolet spectra, the same lines were 
generally used for both stars and, hence, the abundance ratios between the 
stars are independent of the adopted $gf$-values.




Several consistency checks were applied to our analyses, especially to the
more complete analysis of V1920\,Cyg. Ionization equilibrium 
for Fe\,{\sc ii}/Fe\,{\sc iii} is satisfied for both stars,
and for S\,{\sc ii}/S\,{\sc iii} for V1920\,Cyg. Excitation
equilibrium is found for the Fe\,{\sc iii}
optical lines for V1920\,Cyg; the ultraviolet lines used in our analyses
do not offer a range in their lower excitation potential.
When an element provides lines in the optical and the ultraviolet
spectra of V1920\,Cyg, the abundances are in good agreement.
Our results for V1920\,Cyg are in good agreement with those by
Jeffery et al. from a lower resolution optical spectrum of
limited bandpass. The agreement for HD\,124448
 is less good between our ultraviolet-based
abundances and those from a photographic optical spectrum
(Heber 1983). Our limit on the
 H abundance for V1920\,Cyg from the absence of the
H$\alpha$ line is more than 2 dex less than the abundance offered
by Jeffery et al. from the H$\beta$ line. Heber (1983) put the H abundance of 
HD\,124448 at $\log\epsilon$(H) $<$ 7.5.   

\section{Neutron-capture Elements}

Our optical and ultraviolet spectra of hot EHes were scanned for
lines of Y\,{\sc iii}, Zr\,{\sc iii}, and of the doubly-ionized
lanthanides. 
Inspection of the ultraviolet spectra showed that
the two EHes differ dramatically in the abundances of Y
and Zr. This point is highlighted by Figure 1. In Figure 1a,
the Zr\,{\sc iii} line at 2656.47 \AA\ is comparable in
strength to the Mg\,{\sc ii} lines at 2660.8 \AA\ in V1920\,Cyg
(lower spectrum) but in HD\,124448 the Zr\,{\sc iii}
line is extremely weak. Figure 1b shows a similar,
albeit less dramatic comparison on account of blends, for the
Y\,{\sc iii} line at 2367.2 \AA. These differences are
not attributable to differences in stellar parameters. 
This comparison  eliminates the possibility that
the lines which we attribute to Y\,{\sc iii} and Zr\,{\sc iii} 
are merely unidentified lines of iron or other more 
abundant species. 

The measured wavelengths of Y\,{\sc iii} and Zr\,{\sc iii}
lines were taken from Epstein \& Reader (1975)
and Khan, Chaghtal \& Rahimullah (1981), respectively. The 
Y\,{\sc iii} resonance lines at 2414.60, 2367.23, and
2327.31\AA\ are detected in the spectrum of V1920\,Cyg with the
2367.23\AA\ line seen as a contributor to a blended line in
the spectrum of HD\,124448 (Figure 1). 
Lines of Zr\,{\sc iii} at
2664.27, 2656.47, 2643.82,
2620.56, 2593.70, 2102.26, 1921.94, and 1863.98\AA\ 
in the spectrum of V1920\,Cyg were used to set the
abundance.
 Three of the Zr\,{\sc iii} lines at 2664.27, 2656.47,
and 2643.82\AA\ were detectable in the spectrum of HD\,124448.

Figure 1 shows our syntheses of the region around 
2656.47 \AA.
 A small change in
the assumed Fe abundance between V1920\,Cyg ($\log\epsilon$(Fe) = 6.8)
and HD\,124448 ($\log\epsilon$(Fe) = 7.1) is recognized.

\begin{figure*}[t]
\centerline{\psfig{file=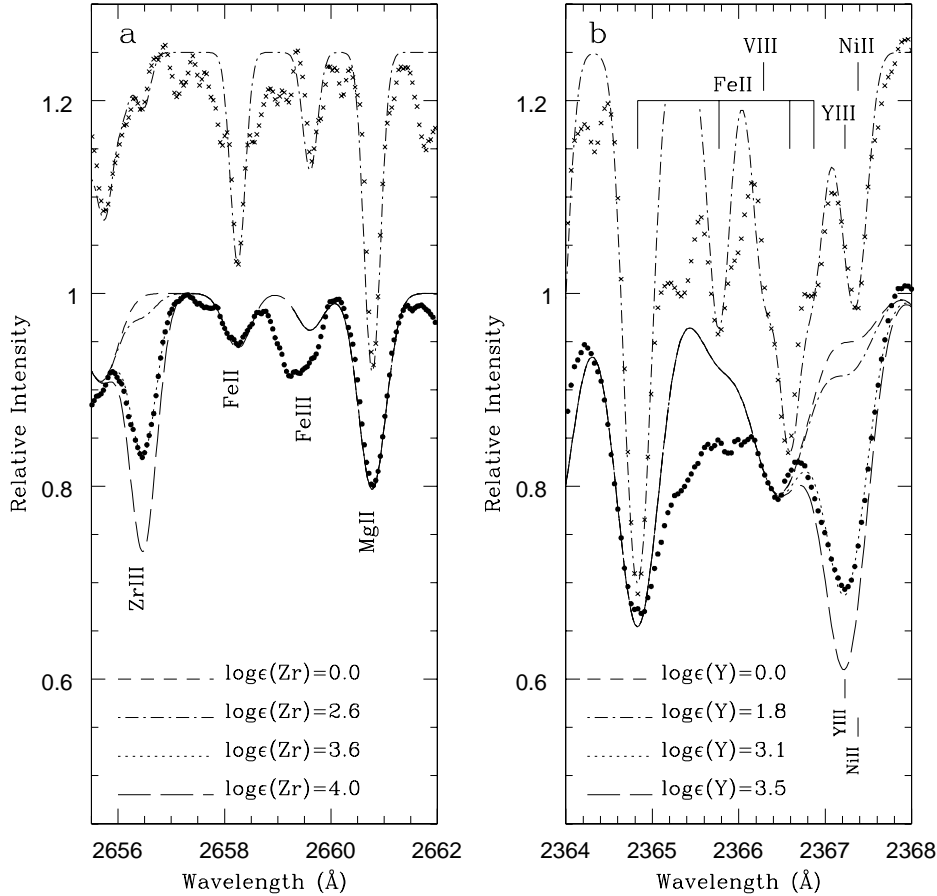,width=13.0cm,angle=0}}
\figcaption[fig1.ps]{
The observed spectra of
V1920\,Cyg and HD\,124448 are represented by filled circles and crosses, respectively.
Panel-a shows the region including the  Zr\,{\sc iii} line
 at 2656.5\AA. Synthetic spectra for four different Zr abundances are shown
for V1920\,Cyg with $\log\epsilon$(Zr) = 3.6 providing a satisfactory fit to
the observed line. The abundance $\log\epsilon$(Zr) = 2.6 provides a fit to
the sharper line in the HD\,124448 spectrum.
Panel-b shows the  region including the Y\,{\sc iii} line  at 2367.2\AA\
which is blended with a Ni\,{\sc ii} line.
The synthetic spectrum with $\log\epsilon$(Y) = 3.1 provides a fit to
V1920\,Cyg's spectrum, and $\log\epsilon$(Y) = 1.8 to HD124448's
spectrum.
In each Panel, principal lines are identified.}
\end{figure*}

An upper limit to the Sr abundance for V1920 Cyg
 is estimated from the non-detection
of the resonance line of Sr\,{\sc ii} 4215.52\AA.

An unsuccessful search was  conducted for  
the doubly-ionized lanthanides. We mention here only those
lanthanides providing significant upper limits to the
abundances. A La\,{\sc iii} 2379.37\AA\ resonance line 
provides the upper limit for V1920\,Cyg. 
An upper limit to the Ce abundance is obtained by comparing the synthetic
and observed profile of the low excitation Ce\,{\sc iii} 2603.59\AA\ line.
The two strongest resonance lines of Nd\,{\sc iii}
at 5193.06\AA\ and at 5294.10\AA\ set the 
upper limit to the Nd abundance in V1920\,Cyg. 


\section{Discussion}

Our abundance analysis  of optical and ultraviolet spectra
of V1920\,Cyg confirms the results obtained by Jeffery et al. (1998)
for elements from C to the iron-group. We extend the earlier analysis to
Mn, Co, Ni, and Zn, and, in particular, to the neutron-capture elements
Y and Zr. We similarly confirm and extend the previous analyses
of HD\,124448 (Sch\"{o}nberner \& Wolf 1974; Heber 1983). We have provided
the first, abundances of neutron-capture elements for hot EHes.
These neutron-capture element abundances
suggest that at least in the EHe star V1920\,Cyg, $s$-process nucleosynthesis 
did occur in its earlier evolution.

An interesting similarity is suggested by these abundances and those of the
R\,CrBs.
A feature of the R\,CrB stars is the large
star-to-star variation 
in the abundances of the
neutron-capture elements. For `majority' R\,CrBs (Lambert \& Rao 1994), [Y/Fe] and [Zr/Fe]
ranges from +0.3 to about +1.6 but with smaller values for [Ba/Fe],
[La/Fe], and presumably other lanthanides (Asplund et al. 2000). This
range and the non-solar ratio of Y and Zr to Ba and lanthanides
was confirmed by Rao \& Lambert (2003) from a differential analysis of the 
newly discovered  R\,CrB star V2552 Oph and R\,CrB. Among 
`minority' R\,CrBs (i.e., very Fe-poor stars), somewhat
more extreme values are known with a similar contrast
between light and neutron-capture elements.
The cool peculiar R\,CrB U Aqr has very extreme neutron-capture element
enrichments: [Y/Fe] = +3.3 and [Ba/Fe] = +2.1 (Vanture, Zucker \&
Wallerstein 1999; also Bond, Luck \& Newman 1979).  
Limited data for cool EHes suggest
Sr, Y, or Zr abundances within the R\,CrB range, say [X/Fe] of 0.0 to
+0.9 (Pandey et al. 2001). 

Our STIS spectra demonstrate that the star-to-star
variation in [Y/Fe] and [Zr/Fe] among hot EHes is at least
as great as among the `majority' R\,CrB stars: V1920\,Cyg with 
[Y/Fe] $\simeq$ [Zr/Fe] $\simeq$ 1.6 is at  one end of the range and  
HD\,124448 with [Y/Fe] $\simeq$ [Zr/Fe] $\simeq$ +0.1 is at the other end. 
Unfortunately, the upper limits set on abundances of lanthanides do not permit
us to  check that EHes follow R\,CrBs in showing smaller
enrichments of these heavier elements. The upper limits set for
Ce and Nd in V1920\,Cyg ([Ce/Fe] $\leq$+1.1 and [Nd/Fe] $\leq$+1.0)
hint at a behavior similar to the R\,CrBs.

As noted in the Introduction, two scenarios are potential sources of
H-deficient luminous stars. A final-flash in a post-AGB stars, and
a merger of a He with a C-O white dwarf. The C and O abundances
of the hot EHes are consistent with predictions of white dwarf mergers but 
not the current final-flash models (Pandey et al. 2001; Saio \& Jeffery 2002).
Saio \& Jeffery (2002) also show that
a EHe star formed by accretion of He-rich material by a C-O white
dwarf has the pulsational properties of real EHe stars. Additionally,
a merger better accounts for large line widths seen in spectra of EHes;
the accreting star is spun up by the accreted gas from the (former) orbiting
He white dwarf.
Yet, the neutron-capture element abundances of the EHes are, perhaps, more
readily explained by
the final-flash scenario, as suggested by the final-flash
candidates FG Sge and V4334\,Sgr with neutron-capture element overabundances.
The range in relative
overabundances of Y and Zr to Ba and the lanthanides varies greatly
between the pair:
 V4334\,Sgr
 has a high relative overabundance (Asplund et al. 1997),
but FG\,Sge has a low relative 
overabundance (Gonzalez et al. 1998).

Enrichment of neutron-capture elements is not expected for the EHes
unless synthesis by neutrons via the $s$-process occurs during the merger
(Pandey et al. 2001). 
The surface of the C-O white dwarf, the former core of
an AGB star, will  be rich in $s$-process products but little
of this C-rich material is required at the surface of the star after
accretion of He-rich material to account for the observed abundances
of the light elements.
 A thin residual He-shell around the C-O white dwarf
core could  contribute $s$-process products.
 The He white dwarf and its possible H-rich skin  are not expected to be
rich in neutron-capture elements.
 One supposes that accretion of the He white dwarf 
may be accompanied by an episode of nucleosynthesis including
 release of neutrons through $^{13}$C$(\alpha,n)^{16}$O
with $^{13}$C created by H-burning. 
If, as seems plausible,   the strength of the neutron
source and efficiency of neutron captures
 varies from merger to merger,  
the range in the neutron-capture element abundances, as
seen here for the EHe pair V1920\,Cyg and HD\,124448 and
known among R\,CrBs, results. Aspects of the composition of
the minority R\,CrBs suggest nucleosynthesis may accompany
the accretion  of He-rich material by the C-O white dwarf. 
Theoretical studies of the merger scenario are to be sought with
careful examination of the attendant nucleosynthesis.

We thank the referee Glenn Wahlgren for an incisive report,
 Carlos Allende Prieto for reading and commenting on a draft of
this paper. We acknowledge support from the Space Telescope
Science Institute through grant GO-09417.


\begin{thebibliography}{}

\bibitem[]{}
Asplund, M., Gustafsson, B., Lambert, D. L., \& Rao, N. K., 1997, A\&A, 321, L17
\bibitem[]{}
Asplund, M., Gustafsson, B., Lambert, D. L., \& Rao, N. K., 2000, A\&A, 353, 287
\bibitem[]{}
Bi\'{e}mont, E., Quinet, P., \& Ryabchikova, T. A., 2002, MNRAS, 336, 1155.
\bibitem[]{}
Bond, H. E., Luck, R. E., \& Newman, M. J., 1979, 233, 205
\bibitem[]{}
Charro, E., L\'{o}pez-Ayuso, J. L., \& Mart\'{i}n, I., 1999, J. Phys. B, 32, 4555
\bibitem[]{}
Duerbeck, H. W., \& Benetti, S., 1996, ApJ, 468, L111
\bibitem[]{}
Epstein, G. L., \& Reader, J., 1975, J. Opt. Soc. Am. A65, 310
\bibitem[]{}
Gonzalez, G., Lambert, D. L., Wallerstein, G., Rao, N. K., Smith, V. V., \&
McCarthy, J. K., 1998, ApJS, 114, 133.
\bibitem[]{}
Heber, U., 1983, A\&A, 118, 39
\bibitem[]{}
Iben, I. Jr., Kaler, J. B., Truran, J. W., \& Renzini, A., 1983, ApJ, 264, 605
\bibitem[]{}
Jeffery, C. S., Drilling, J. S., \& Heber, U., 1987, MNRAS, 226, 317
\bibitem[]{}
Jeffery, C. S., Hamill, P. J., Harrison, P. M., \& Jeffers, S. V., 1998, A\&A, 340, 476
\bibitem[]{}
Jeffery, C. S., Woolf, V. M., Pollacco, D. L. 2001, A\&A, 376, 497
\bibitem[]{}
Khan, Z. A., Chaghtal, M. S. Z., \& Rahimullah, K., 1981, Phys. Scripta, 23, 29
\bibitem[]{}
Lambert, D. L., \& Rao, N. K., 1994, JAA, 15, 47
\bibitem[]{}
Langer, G. E., Kraft, R. P., \& Anderson, K. S., 1974, ApJ, 189, 509
\bibitem[]{}
Lodders, K., 2003, ApJ, 591, 1220
\bibitem[]{}
Pandey, G., Rao, N. K., Lambert, D. L., Jeffery, C. S., \& Asplund, M., 2001, MNRAS, 324, 937
\bibitem[]{}
Rao, N. K., \& Lambert, D. L., 2003, PASP, 115, 1304
\bibitem[]{}
Reader, A., \& Acquista, N., 1997, Phys. Scripta, 55, 310
\bibitem[]{}
Redfors, A., 1991, A\&A, 249, 589
\bibitem[]{}
Saio, H., \& Jeffery, C. S., 2002, MNRAS, 333, 121
\bibitem[]{}
Sch\"{o}nberner, D., \& Wolf, R. E. A., 1974, A\&A, 37, 87
\bibitem[]{}
Sikstr\"{o}m, Lundberg, H., Wahlgren, G. M., Li, Z. S., Lynga, C., Johansson, S., \& Leckrone, D. S., 
    1999, A\&A, 343, 297
\bibitem[]{}
Tull, R. G., MacQueen, P. J., Sneden, C., \& Lambert, D. L., 1995, PASP,
    107, 251
\bibitem[]{}
Vanture, A. D., Zucker, D., \& Wallerstein, G., 1999, ApJ, 514, 932
\bibitem[]{}
Webbink, R. F., 1984, ApJ, 277, 355
\bibitem[]{}
Wiese, W. L., Fuhr, J. R., \& Deters, T. M., 1996, Journal of Physical and Chemical Reference Data, Monograph No. 7
\bibitem[]{}
Zhang, Z. G., Svanberg, S., Palmeri, P., Quinet, P., \& Bi\'{e}mont, E., 2002, A\&A, 385, 724


\end{thebibliography}
\end{document}